\documentclass[conference]{IEEEtran}
\IEEEoverridecommandlockouts
\usepackage{cite}
\usepackage{amsmath,amssymb,amsfonts}
\usepackage{algorithmic}
\usepackage{graphicx}
\usepackage{rotating}
\usepackage{tikz}
\usepackage{subfig}
\usepackage{textcomp}
\usepackage{xcolor}
\usepackage{url}
\usepackage{multirow}
\usepackage{tabularx} 

\def\BibTeX{{\rm B\kern-.05em{\sc i\kern-.025em b}\kern-.08em
    T\kern-.1667em\lower.7ex\hbox{E}\kern-.125emX}}

\usepackage{color, colortbl}
\definecolor{Gray}{gray}{0.8}

\usepackage{fancyhdr}
\begin{document}

\title{Knowing Your Annotator:\\Rapidly Testing the Reliability of Affect Annotation

\thanks{This project has received funding from the European Union’s Horizon 2020 programme under grant agreement No 951911.
Antonios Liapis and Georgios N. Yannakakis were supported by the Malta Council for Science and Technology (MCST) under the FUSION R\&I: Research Excellence programme (Project number: REP-2022-017).
}
}

\author{\IEEEauthorblockN{
Matthew Barthet,
Chintan Trivedi,
Kosmas Pinitas,
Emmanouil Xylakis, \\
Konstantinos Makantasis,
Antonios Liapis,
Georgios N. Yannakakis\\\\
\IEEEauthorblockA{Institute of Digital Games, University of Malta, Msida, Malta.\\
Email: \{matthew.barthet, ctriv01, kosmas.pinitas, emmanouil.xylakis, \\ konstantinos.makantasis, antonios.liapis, georgios.yannakakis\}@um.edu.mt}}}

\maketitle
\thispagestyle{fancy}

\begin{abstract}
The laborious and costly nature of affect annotation is a key detrimental factor for obtaining large scale corpora with valid and reliable affect labels. Motivated by the lack of tools that can effectively determine an annotator's reliability, this paper proposes general quality assurance (QA) tests for real-time continuous annotation tasks. Assuming that the annotation tasks rely on stimuli with audiovisual components, such as videos, we propose and evaluate two QA tests: a visual and an auditory QA test. We validate the QA tool across 20 annotators that are asked to go through the test followed by a lengthy task of annotating the engagement of gameplay videos. Our findings suggest that the proposed QA tool reveals, unsurprisingly, that trained annotators are more reliable than the best of untrained crowdworkers we could employ. Importantly, the QA tool introduced can predict effectively the reliability of an affect annotator with $80\%$ accuracy, thereby, saving on resources, effort and cost, and maximizing the reliability of labels solicited in affective corpora. The introduced QA tool is available and accessible through the PAGAN annotation platform.
\end{abstract}


\section{Introduction} \label{sec:introduction}
Revealing and capturing the \textit{ground truth of affect} is arguably the fundamental challenge of affective computing. Over the 20 or more years of the existence of the field, several approaches have been introduced as a response to this challenge. Generally speaking, those vary between tools that attempt to minimize inter-rater agreement (e.g.\cite{feeltrace, gtrace, lopes2017ranktrace, yannakakis2015grounding}) and data processing methods that attempt to derive the ground truth from any affect labels provided \cite{yannakakis2017ordinal,yannakakis2018ordinal,booth2020fifty}. Among the several factors that determine the reliability of affect annotation, of key importance is the \textit{reliability of the annotator} per se. This can be caused by lack of understanding of the task or annotation labels, inattentiveness, poor or faulty equipment, and many more uncontrollable parameters when dealing with in-the-wild settings. In human-computer interaction tasks, the potential of any tool is often established through Quality Assurance (QA) tests, since the intended quality is described by the tool's designer and can be assessed. However, in affective computing it is not trivial to assess the quality of a completed annotation task (e.g. a trace or a label) as the ground truth of affect is not available; it is derived from the annotators themselves. Hence measuring any deviation from an assumed (and likely moving) goal is ill-posed by definition. 

Motivated by the lack of tools for assessing the quality of affect annotators, this paper introduces a set of accessible QA tests that can determine the reliability of an annotator rapidly and effectively. We focus on audiovisual content annotation and we use RankTrace \cite{lopes2017ranktrace} from the PAGAN \cite{PAGAN} suite of continuous annotation tools. To assess the reliability of annotators on an audiovisual task, our QA tools require annotators to first annotate two videos containing an objective task with a predetermined ground truth. Each video has a duration of 1 minute and tests the annotators' ability to annotate visual stimuli (visual QA test) and aural stimuli (auditory QA test). 

The QA tests introduced are tested in a real-world use-case regarding the time-continuous annotation of engagement in gameplay videos. To test these QA tools, we employ and compare two groups of annotators with different levels of annotation expertise. The first group consists of 10 trained affective computing researchers from the University of Malta. The second group are 10 untrained crowdworkers employed from Amazon's Mechanical Turk (MTurk). In the protocol reported in this paper, we first ask participants to go through our two introduced QA tests. We then ask them to annotate the engagement levels of 30 gameplay videos. Based on the data obtained, we test three hypotheses: (a) that the trained annotators are more reliable than the untrained crowdworkers; (b) that for audiovisual annotation tasks it is necessary to assure the reliability of annotators through both auditory and visual QA tasks; (c) that the performance of annotators on the objectively-defined tests can be a reliable predictor of their reliability in their followup affect annotation tasks.

Our findings validate all hypotheses and suggest that (a) annotator training is critical for obtaining reliable labels of affect; (b) each of the two QA tasks can reveal unreliable annotators which may or may not overlap; and (c) that short and trivial annotator tests can be used as accurate predictors of affect annotator quality with up to $80\%$ accuracy. The benefit of the proposed tools is that they are easily deployed by researchers and rapidly executed by annotators, thereby reducing the costly and laborious nature of annotation. The QA tools can likely detect unreliable annotators, which can increase the overall reliability of the affect labels obtained. The introduced annotator reliability tests are accessible through the training \cite{PAGAN} tool for the rapid assessment of participants of any audiovisual affect annotation task.

\section{Background: Affect Annotation Reliability} \label{sec:background}

Arguably one of the most crucial steps in any data labeling process is the choice of the annotation task, which, in turn, highly affects the quality of the solicited dataset. The task selection process is relatively straightforward when the annotations can be defined \emph{objectively} (e.g. a cat vs a dog class label) since a single best response is presumed to exist for each sample. However, when it comes to \emph{subjective} annotation tasks such as arousal or valence traces, the ground-truth signal is ill-posed due to inherent subjectivity bias of annotators and their potential systematic reporting errors \cite{yannakakis2017ordinal,yannakakis2018ordinal}.

Due to the subjective nature of emotion, several studies in affective computing focus on the assessment of a label's validity and reliability \cite{booth2017toward,mckeown2011semaine,metallinou2013annotation}. The validity of an affect label can be defined as the degree to which the annotation measures the phenomenon we claim it does and it can be quantified using cross-validation \cite{yannakakis2017ordinal} or physiological measures \cite{kuster2017measuring}. The reliability of an affect label instead (the focus of this paper) is usually quantified through variant measures of inter-rater agreement.

As noted above, the absence of an objective ground truth signal for comparison renders the assessment of a human annotator's reliability a notoriously difficult challenge. In the literature, the dominant method for determining the quality of an annotator is to compare their affect traces with those of other annotators on the same task using a number of inter-rater agreement measures (see Section \ref{sec:results_metrics}). Indicatively, Aljanaki \emph{et al.} \cite{aljanaki2017developing} employed Cronbach's $\alpha$ on crowd-sourced affect labels to remove annotations that fall out of the majority cluster. Kossaifi \emph{et al.} \cite{kossaifi2017afew}, instead, use Pearson's correlation to improve the reliability of the annotations obtained by removing low-agreement annotations. Yannakakis \emph{et al.} \cite{yannakakis2015grounding} measured inter-annotator agreement for ordinal and interval-based human emotion labels by employing Krippendorff's $\alpha$ whereas Devillers \emph{et al.} \cite{devillers2006real} used Cohen’s $\kappa$ coefficient along with a thresholding scheme forming a consensus among annotators and discarding outliers. Nicolaou \emph{et al.}  \cite{nicolaou2010automatic} introduced the signed agreement metric for evaluating the accordance of continuous valence annotations. Expanding on this, Booth and Narayanan \cite{booth2020fifty} introduced an ordinal measure of agreement, \emph{signed differential agreement} (SDA), and examined several agreement metrics in a controlled study with a predetermined and known ground truth.

Similarly, selecting the right agreement measure is a challenging endeavor. This is especially true in continuous and unbounded affect signals, such as those generated by RankTrace \cite{lopes2017ranktrace}, as annotators will likely leave a trace that embeds their own subjective and reporting biases. Building on recent developments in ordinal affect annotation \cite{yannakakis2017ordinal, yannakakis2018ordinal,booth2020fifty,metallinou2013annotation} we argue that agreement measures for affect traces should consider the annotator's consensus in terms of the trace's relative changes instead of their agreement in terms of the trace's absolute values. In contrast to the aforementioned studies, in this paper we compare the performance of trained researchers against untrained crowdworkers across several reliability measures and examine the capacity of our introduced tools to predict the reliability of an annotator.

\begin{figure}[!tb]
\centering
\subfloat[Visual QA Task]{
\includegraphics[width=.235\textwidth]{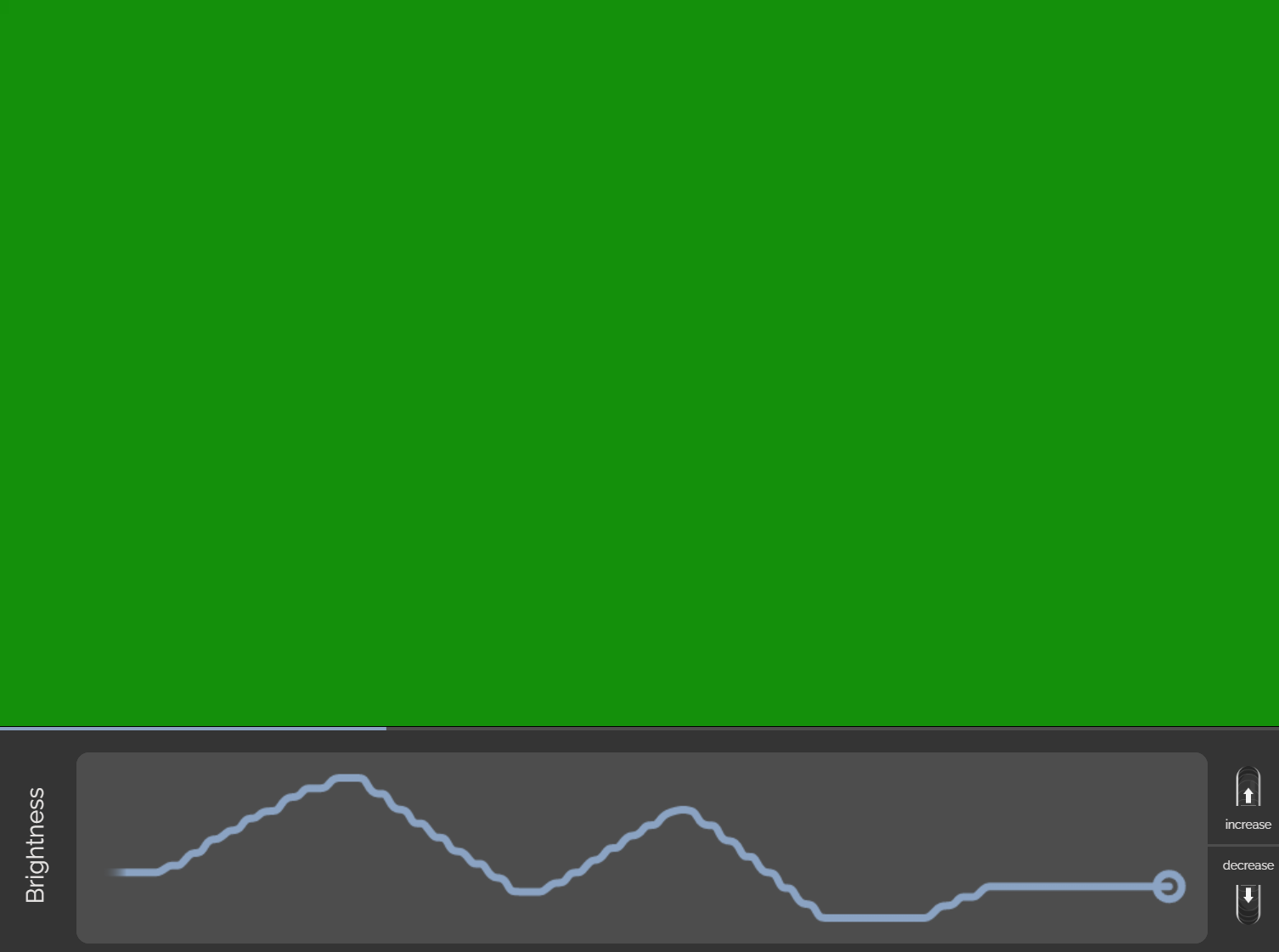}\label{fig:ui_visual}
}
\subfloat[Auditory QA Task]{
\includegraphics[width=.235\textwidth]{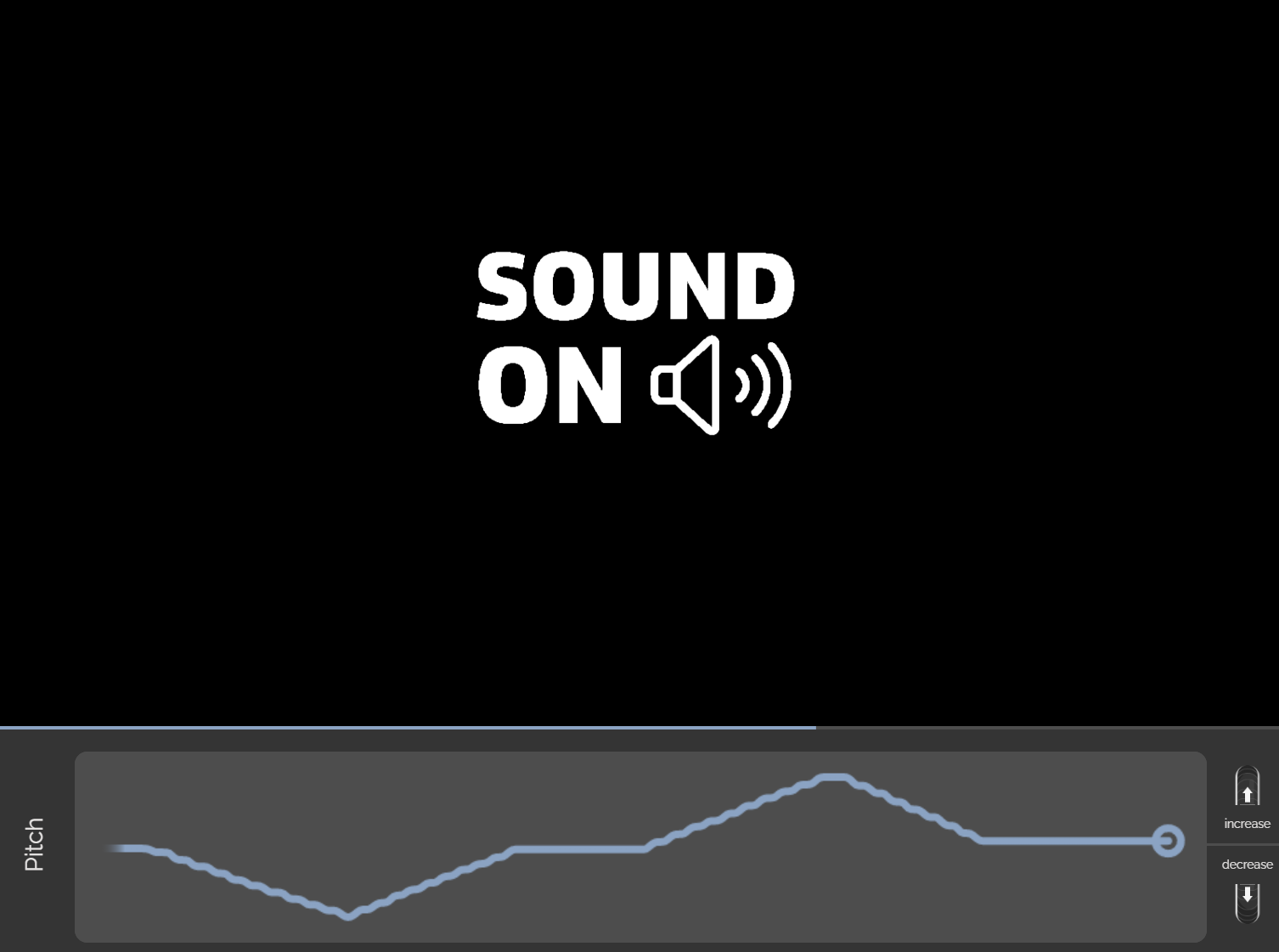}\label{fig:ui_auditory}
}
\caption{User interface for the QA tests used in this study. The two tests introduced utilize RankTrace \cite{lopes2017ranktrace} and are implemented through the PAGAN framework \cite{PAGAN}.}
\label{fig:PAGAN}
\end{figure}

\section{Quality Assurance Tests For Audiovisual Annotation Tasks}\label{sec:qa}

Below we outline the two Quality Assurance (QA) tools we introduce for testing the reliability of annotators (see Fig. \ref{fig:PAGAN}): one visual QA test and one auditory QA test. The QA tests are founded on the premise that the audio or visual content used as stimuli are (a) simple to annotate in terms of their labels, and (b) fully controlled by the researchers to provide an \emph{objective ground truth}. The stimuli in both QA tests listed below are in video format, and are created within the Processing environment\footnote{https://processing.org/}. The videos, ground truth signal, and instructions on integrating these with PAGAN \cite{PAGAN} are available online\footnote{\url{https://github.com/institutedigitalgames/pagan_qa_tool_sources}} for researchers interested in time-continuous annotation tasks.

The QA videos are uploaded and made available through the online PAGAN annotation interface \cite{PAGAN}, following the RankTrace annotation protocol. In RankTrace, participants can modify their perceived affect assessment in real-time (through the scroll-wheel of the mouse) while they watch a video recording and reacting to stimuli as they occur. The entire annotation trace is visible to the participant through the interface (bottom of the screen in Fig.~\ref{fig:ui_visual}), scaling it on the $x$-axis as the video continues. The resulting annotation traces collected via PAGAN are time-continuous and unbound, which we process through min-max normalization for the purposes of this paper (see Fig.~\ref{fig:all_traces}).

We argue that the use of these QA tests is beneficial for minimizing computational and human (annotator) resources of any affect annotation task while maximizing the reliability of the affect labels obtained. Our hypotheses are supported by the findings of this paper, detailed in Section \ref{sec:evaluation}. 

\subsection{Visual QA Test}\label{sec:qa_visual} 

Inspired by \cite{booth2020fifty}, our first test in this QA toolkit (see Fig.~\ref{fig:ui_visual}) consists of a purely visual task where the annotator is requested to annotate the change of brightness of a green screen. Before the test starts, the user is prompted with the following instructions: \emph{``Please use the scroll-wheel to indicate the changes in the level of brightness while watching the video''}. The annotator uses a mouse wheel to scroll up or down to indicate the change of color brightness observed. 

The 1-minute video used as a stimulus has a predetermined pattern of brightness changes. The green color intensity value ranges between 25 and 255, and its fluctuation changes are controlled by the script that produces the video. The min-max normalized fluctuations of green color intensity are shown in the solid black line of Fig. \ref{fig:trace_qa_vis_expert}. The red and blue channels remain static at values of 20 and 12, respectively. This results in frames where the screen is dark and progressively more or less green, avoiding issues in perceptible brightness due the RGB file format.  We carefully selected this test as its objective nature and low cognitive load allows us to effectively test the annotator's ability to respond to visual stimuli and their response time.

\subsection{Auditory QA Test}\label{sec:qa_auditory}

Based on the premise of the visual QA test, and due to the nature of audiovisual stimuli used in the actual annotation task (see Section \ref{sec:data_collection}), we present the annotator with an auditory QA test (see Fig.~\ref{fig:ui_auditory}). The 1-minute video used as a stimulus for this task has no visual component (showing a black screen with a message, as in Fig.~\ref{fig:ui_auditory}) and a monotonic audio signal with a dynamically changing pitch. The pitch ranges between 50 and 470 Hz, and the sound is produced by a triangle wave oscillator. As with the visual task, the ground truth of the pitch variation is normalized and illustrated as a black solid line in Fig. \ref{fig:trace_qa_aud_expert}. Once again, the annotator is requested to use RankTrace \cite{lopes2017ranktrace} and is prompted with the following instructions before the QA task: \emph{``Please use the scroll-wheel to indicate the changes in the level of Pitch while watching the video''}.

This task is designed to test the annotator's ability to react to purely aural stimuli, and to ensure that the annotator is using the proper setup to detect even low-pitch sounds heard in a video. We argue that these two tests, combined, cover both major types of audiovisual stimuli during the annotation process. Testing both types in isolation allows us to ensure the annotator's reliability on both auditory and visual stimuli independently.

\section{Use Case: Annotation of Engagement in Audiovisual Game Stimuli}\label{sec:data}

\begin{figure}
\centering
\includegraphics[width=\columnwidth]{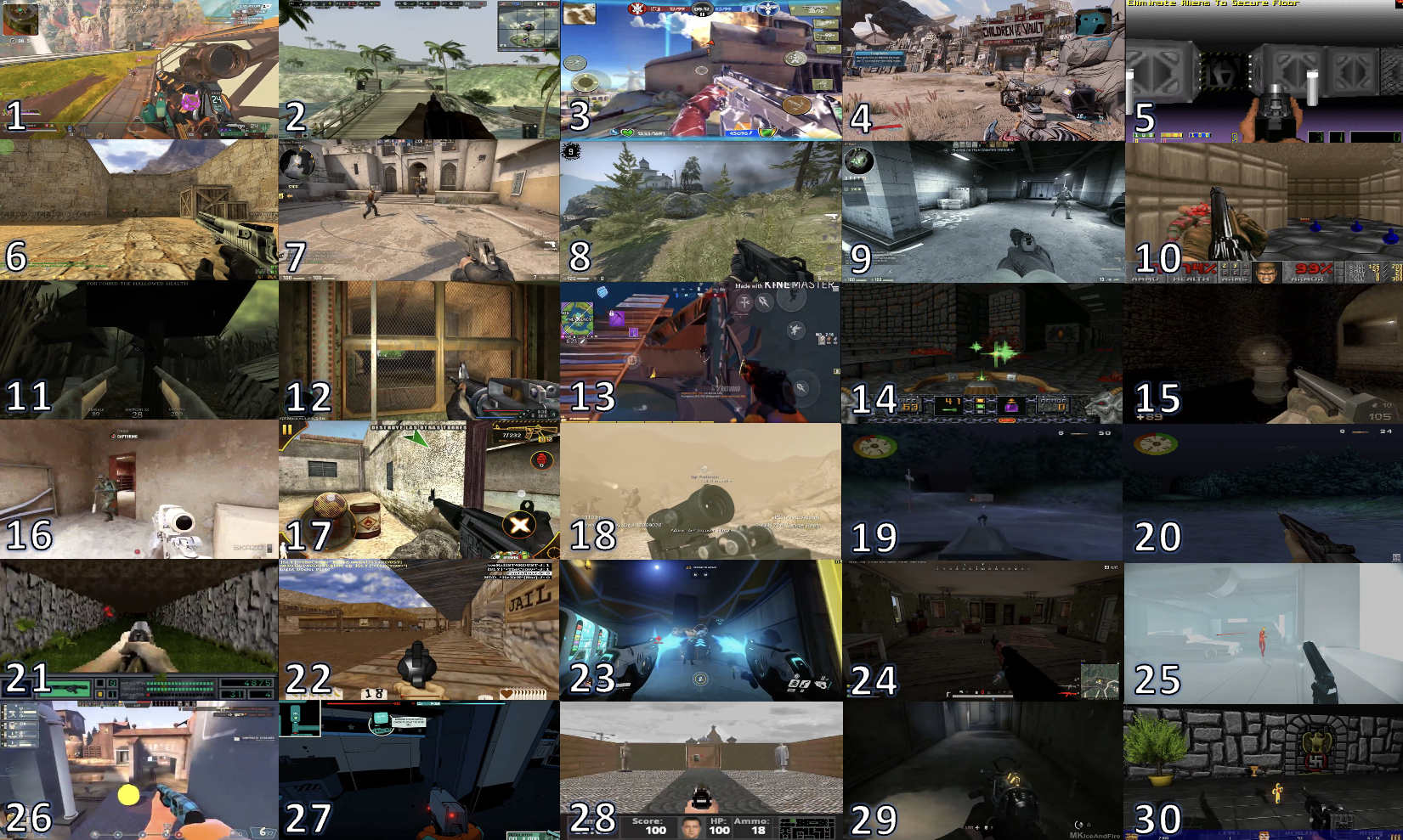}
\caption{Screenshots from the 30 different FPS games annotated for engagement in this paper. List of game titles: (1) Apex Legends; (2) Battlefield 1942; (3) Blitz Brigade; (4) Borderlands 3; (5) Corridor 7; (6) Counter Strike 2016; (7) Counter Strike 2018; (8) Counter Strike 2019; (9) Counter Strike Go; (10) Doom; (11) Dusk; (12) Far Cry 1; (13) Fortnite; (14) Heretic; (15) Hrot; (16) Insurgency; (17) Modern Combat: Sandstorm; (18) Medal of Honor 2010; (19) Medal of Honor 1999; (20) Medal of Honor: Pacific Assault; (21) Operation Bodycount; (22) Outlaws; (23) Overwatch 2; (24) PUBG; (25) Superhot; (26) Team Fortress 2; (27) Void Bastards; (28) Wolfenstein 3D; (29) Wolfenstein New Order; (30) Wolfram Wolfenstein.}
\label{fig:games}
\end{figure}

To evaluate the QA tools described in Section \ref{sec:qa}, we employ a real-world annotation task relevant to our ongoing research goals. It is worth noting that the QA tools were developed precisely to address challenges we encountered during this use case, rather than the other way around. The goal of this annotation task is to produce reliable time-continuous traces of a player's engagement while playing a first-person shooter (FPS) game via third-person annotation. We discuss the stimuli (gameplay videos) in Section \ref{sec:method-videos} and the data collection protocol in Section \ref{sec:data_collection}.

\subsection{FPS Game Videos} \label{sec:method-videos} 

For the purposes of this study we selected gameplay videos from 30 different and popular FPS games (see Fig. \ref{fig:games} for the full list of games) as elicitors of engagement. The choice of FPS games and videos was based on several criteria. First, we wish to ensure that these games cover a wide range of audiovisual stimuli for engagement annotation varying in graphical style (i.e. photo-realistic, retro, cartoon-like, etc.) and gameplay modes (i.e. battle royale, campaign, deathmatch, etc.). Second, we ensure there is no player or user commentary present in the video and that only the sound of the game can be heard during the video playback. Finally, we ensure that there is no video with more than 15 seconds of non-gameplay footage (e.g. menu screens, cut scenes or transition animations). The resulting game videos always have a duration of 1 minute.

In order to provide annotators with a variety of stimuli, each \textit{session} had 30 gameplay videos (one per FPS game) shown to the players in sequence. \textit{Session 1} included therefore 30 videos in the form of a 30-minute sequential annotation task (with videos in random order, see Section \ref{sec:data_collection}). \textit{Session 2} included a different set of 30 gameplay videos (from the same 30 games shown in Fig.~\ref{fig:games}). We consider videos from Session 2 as independent stimuli from those of Session 1, despite originating from the same game, since the events depicted (and thus the perceived player engagement) are different---and differently timed---between gameplay videos from the same FPS game, or similar games. With two sessions of independent stimuli (Session 1 and 2), we produce a corpus of gameplay videos of a total duration of 60 minutes (i.e. 30 minutes for each session). 

\subsection{Data Collection Protocol} \label{sec:data_collection}

To accomplish our overarching goal of annotating gameplay videos in terms of engagement, we implemented the following protocol for both experts and crowdworkers. 

\subsubsection{Participants}

\emph{Experts} were members of the University of Malta (either research staff or M.Sc. students), with a total of 10 experts reported in this paper. All experts performed the annotation in the same room and light conditions, using the same machine and input/output devices (screen for visual stimuli, headphones for aural stimuli, and a mouse with a scroll wheel for annotation). Researchers involved in this work (all of them authors of this paper) were always present to introduce the annotation task and answer any questions during the annotation period. To gather richer data on the experience, expert annotators answered some questions and gave their feedback after the task was completed; this may be processed and reported in future work. 
On the other hand, \emph{crowdworkers} were recruited through Amazon Mechanical Turk (MTurk), and were required to use headphones and a mouse with a scroll wheel to take part in the experiment. Crowdworkers were redirected to the same online PAGAN website as experts (more on this below) but performed the annotation task at their own pace, a location and time of their choice, and using their own equipment.
A total of 25 crowdworkers were recruited to annotate a variety of sessions (each totalling a 30-minute annotation task) but only the 10 most reliable crowdworker annotators were retained after data cleanup due to missing annotations and other technical issues. The 10 crowdworkers selected are the ones yielding the highest reliability---measured through the average signed differential agreement \cite{booth2020fifty}---during the two QA tests. Five of the experts and five of the crowdworkers annotated videos from Session 1 and the remaining 5 experts and 5 crowdworkers annotated videos from Session 2. 

\subsubsection{Annotation Process}
Each participant was allocated a Session (1 or 2) and first performed the visual QA test (Section \ref{sec:qa_visual}). followed by the auditory QA test, as described in Section \ref{sec:qa_auditory}. All participants performed the two QA tasks in the same order. After the completion of the QA tests, participants were provided the following definition of engagement: \emph{``A high level of engagement is associated with a feeling of tension, excitement, and readiness. A low level of engagement is associated with boredom, low interest, and disassociation with the game.''}. Participants were then asked to annotate 30 gameplay videos from the FPS corpus described in Section \ref{sec:method-videos}. Unlike the QA tests, the order of these 30 videos was randomized for each participant. The random order was imposed to minimize participants' habituation effects. The randomization of stimuli order, however, poses a number of challenges in deriving a ground truth on the same stimulus between participants since each video may have been seen early on for one participant (i.e. where the annotation task may still be hard to grasp) or late (i.e. where user fatigue may occur). Participants could pause the video, which stopped the annotation process, and they could opt to start the next video once the previous was completed. Once the engagement annotation was completed (i.e. a task lasting approximately 30 minutes per participant), the participant was thanked for their participation and exited. 

\section{Hypotheses and Reliability Measures}\label{sec:protocol} 
Using the affect annotation task discussed in Section \ref{sec:data}, we collected 30 engagement traces per participant, as well as traces for the two QA tests (visual and audio). Based on the designed experimental protocol we attempt to validate the following three hypotheses (H):
\begin{itemize}
    \item[H1] Experts offer more reliable annotation data than crowdworkers.
    \item[H2] Two QA tests are necessary to assess the reliability of annotators in an audiovisual annotation task.
    \item[H3] The QA tests can early-detect unreliable annotators prior to any affect annotation task.
\end{itemize}

To test our hypotheses, we require good approximations of the ground truth and appropriate measures of inter-rater reliability. Section \ref{sec:results_groundtruth} outlines the method we followed to derive the ground truth from the obtained traces. The ground truth enable us to measure our annotators' reliability via the metrics presented in Section \ref{sec:results_metrics}.

\begin{figure*}
\centering
\subfloat[Visual QA Task (Experts)]{
\includegraphics[width=0.315\textwidth]{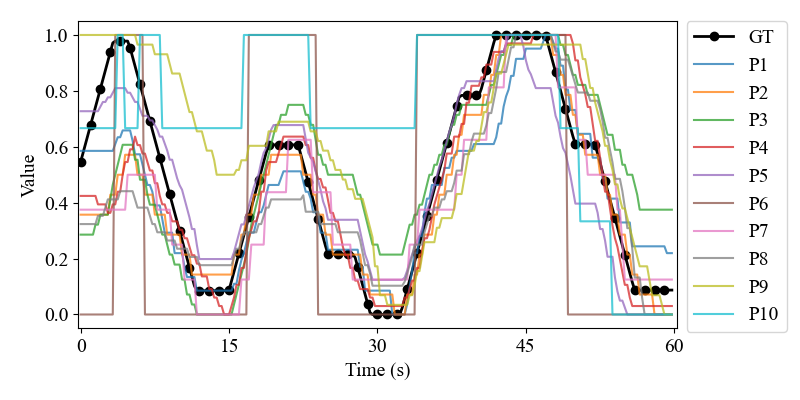}
\label{fig:trace_qa_vis_expert}
}
\subfloat[Auditory QA Task (Experts)]{
\includegraphics[width=0.315\textwidth]{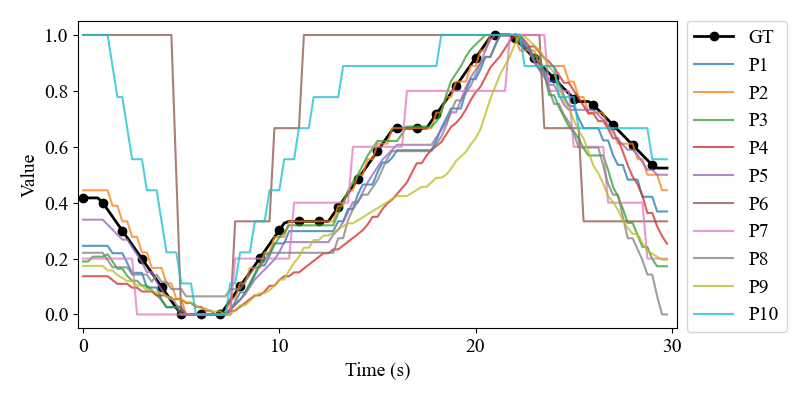}
\label{fig:trace_qa_aud_expert}
}
\subfloat[Engagement Traces (Experts)]{
\includegraphics[width=0.315\textwidth]{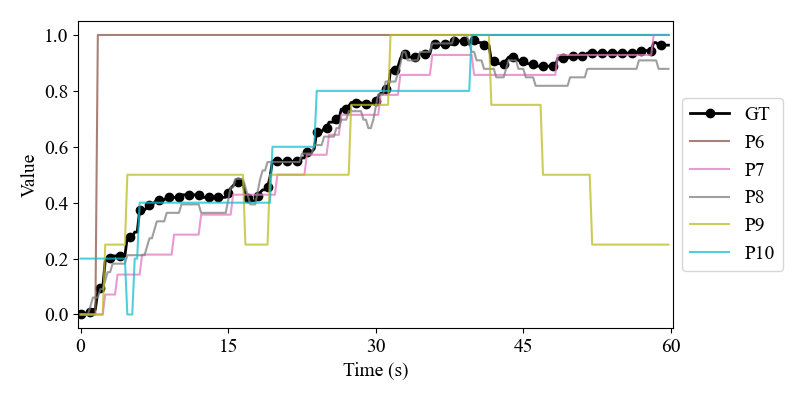}
\label{fig:trace_eng_expert}
}
\\
\subfloat[Visual QA Task (Crowdworkers)]{
\includegraphics[width=0.315\textwidth]{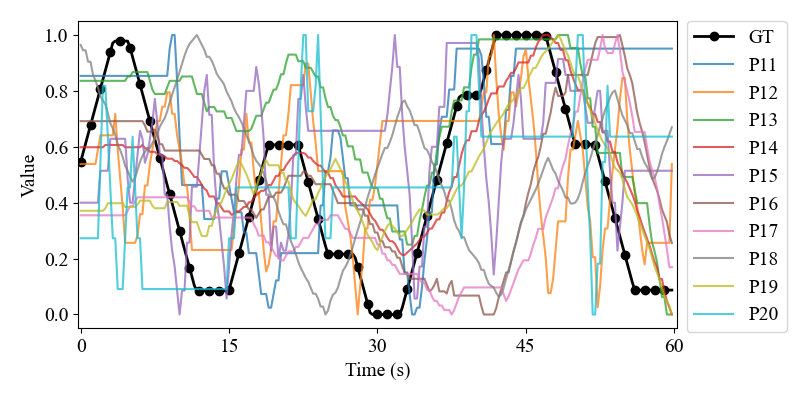}
\label{fig:trace_qa_vis_crowd}
}
\subfloat[Auditory QA Task (Crowdworkers)]{
\includegraphics[width=0.315\textwidth]{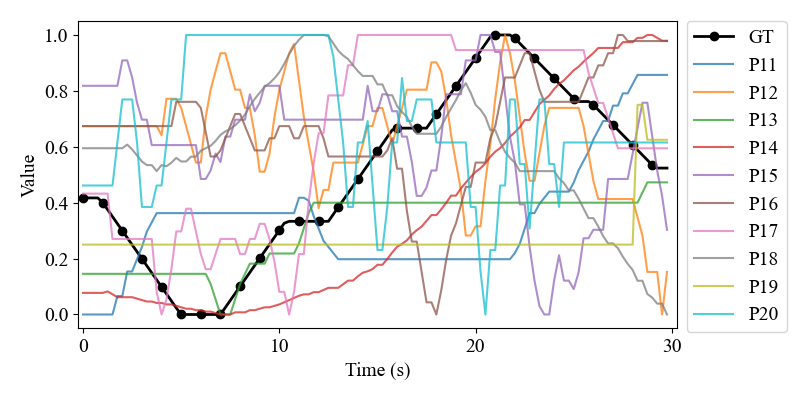}
\label{fig:trace_qa_aud_crowd}
}
\subfloat[Engagement Traces (Crowdworkers)]{
\includegraphics[width=0.315\textwidth]{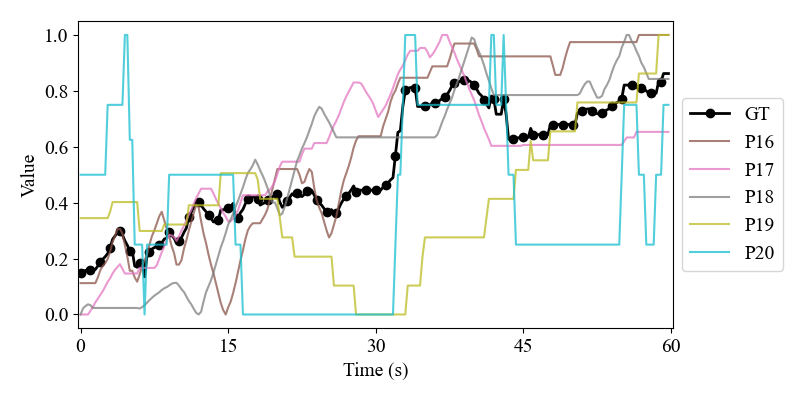}
\label{fig:trace_eng_crowd}
}
\caption{Annotation traces (various colored lines) for the expert (top) and crowdworker (bottom) annotators for the two QA tasks (including all participants) and for an indicative use-case of engagement annotation with the Apex Legends clip in Session 2 (including 5 expert and 5 crowdworker traces). The gold standard signal (GT) is the black solid line, derived from the controlled fluctuations in  stimuli for QA tasks and averaged from the golden standard signals for engagement tasks (see Section \ref{sec:results_groundtruth}).}
\label{fig:all_traces}
\end{figure*}

\subsection{Ground Truth}\label{sec:results_groundtruth}

A strength of the QA tools proposed in this paper is that the ground truth is known in advance: brightness or audio pitch fluctuation is controlled by the researcher \emph{a priori}. Therefore, deriving inter-rater reliability amounts to finding the agreement from this known trace.

While the QA tests come with an objective ground truth trace to compare against the annotators' traces, the gameplay videos do not. Hence, for the corpus under consideration we follow current practices \cite{grewe2007emotions,pinitas2022supervised,yannakakis2018ordinal,makantasis2023lab} and derive a \emph{gold standard signal} (i.e. the median engagement annotation trace) which we consider the ground truth of engagement. We calculate the gold standard signal separately for each session, as they contain different video clips, as well as separately for each group (experts vs crowdworkers). Importantly, when we compare an annotator to their group's gold standard signal, we ensure to leave their data out of the gold standard calculation to eliminate any data leakage. This means that for a group of 5 annotators of the same video, we derive five gold standard signals, one per annotator. This is repeated five times, and we report the average and deviation from those five calculated reliability metrics (one per left-out participant).

\subsection{Reliability Metrics} \label{sec:results_metrics}

Based on a recent analysis of existing agreement metrics \cite{booth2020fifty}, we employ the following statistical measures to calculate annotator reliability in this study: 

\begin{itemize}
\item \textbf{Cronbach's $\pmb\alpha$} \cite{cronbach1951coefficient} is a widely used group measure of internal consistency or reliability of a test or questionnaire. It estimates the extent to which a set of items measures a single, uni-dimensional latent variable, and lies in $[0, 1]$; higher values indicate higher agreement.
\item \textbf{Krippendorff's $\pmb\alpha$} \cite{krippendorff2018content} is a generalization of several inter-rater reliability coefficients, including Cohen's $\kappa$, and is used to measure the agreement between multiple raters who categorize a set of items. It is applicable to any number of raters, levels of measurement (nominal, ordinal, interval, ratio), and incomplete or missing data. Its values range between $[0, 1]$, with higher values indicating higher agreement between raters.
\item \textbf{Cohen's $\pmb\kappa$} \cite{cohen} is a pairwise measure of inter-rater reliability, which assesses the agreement between two raters who independently classify a set of items into discrete categories. It considers the possibility of agreement occurring by chance and lies within $[-1, 1]$, with higher values indicating higher agreement between raters.
\item \textbf{Signed Differential Agreement (SDA)} \cite{booth2020fifty} is a pairwise agreement measure which aims to capture consensus in signal shape and is invariant to perceptual biases. It can be used on both ordinal and interval continuous traces. SDA values lie within $[-1, 1]$ and are computed based on number of times the signs of two signals agree.
\end{itemize}

\section{Results} \label{sec:evaluation} 

In this section we attempt to test the three hypotheses presented in Section \ref{sec:protocol}. To do this, we assess participants' reliability with regards to the known ground truth signals in the QA tests and the derived gold standard signals in the subjective annotation tasks of gameplay engagement (see Section \ref{sec:results_groundtruth}). All reported significance tests are measured at 95\% confidence ($p<0.05$).   

\subsection{H1: Reliability of Experts vs. Crowdworkers} \label{sec:expertsvscrowd}

To test H1, we compare the performance of the 10 expert annotators against that of the 10 crowdworkers split between the two annotation sessions.
Reliability here is measured using all four metrics outlined in Section \ref{sec:results_metrics}: SDA, Cohen's $\kappa$, Cronbach's $\alpha$ and Krippendorff's $\alpha$.

\subsubsection{QA Tests}

Figure \ref{fig:all_traces} shows the normalized traces of all annotators on the visual and auditory tests. It becomes visually apparent that expert annotators---compared to the MTurk workers---not only agree more with each other but they also tend to follow the ground truth much closer. The expert's agreement seems to be strong with regards to both the trace trend (relative change) and the trace's values.  

The visual inspection of Fig. \ref{fig:all_traces} is in alignment with the reliability values obtained across the four measures considered for the two tests (see Table \ref{tab:agreement}). It is clear that the expert group of annotators performs better in terms of reliability across all four measures. Importantly, in terms of SDA, the expert group of annotators have a significantly higher mean SDA than the crowdworkers in both tests. A similar pattern is observed when Cohen's $\kappa$ and Krippendorff's $\alpha$ are used as measures of reliability. The $\kappa$ values of the experts indicate once more that the trained group of annotators is more reliable by a considerable magnitude. Krippendorff's $\alpha$ values reveal the same trend as reliability values for the experts are higher than $0.62$ for both the visual and the auditory tasks whereas they reach levels of no correlation ($\alpha=0$) \cite{krippendorff2018content} among annotators within the group of crowdworkers. While Cronbach's $\alpha$ values indicate a similar trend in both QA tests, both groups achieve very high reliability scores in contrast to the other three metrics. This result is in alignment with the findings of \cite{booth2020fifty} suggesting that Cronbach's $\alpha$ does not capture well the structural similarities of continuous annotations.

\begin{table}
\caption{Annotator reliability across the two QA tests with 95\% confidence intervals for pairwise agreement measures.} \label{tab:agreement}
\begin{center} 
\begin{tabular}{|c|c|c|}
\hline \hline
\multicolumn{3}{|c|}{\textbf{Visual QA Test}} \\
\hline\hline
\textbf{Metric} & \textbf{Experts} & \textbf{Crowdworkers} \\\hline
SDA & $0.09\pm0.15$  & $-0.30\pm0.13$ \\\hline
Cohen's $\kappa$  & $0.27\pm0.14$ & $0.00\pm0.09$\\\hline
Cronbach's $\alpha$ & 0.98 & 0.96  \\\hline
Krippendorff's $\alpha$ & 0.62 & 0.16 \\\hline
\hline 
\multicolumn{3}{|c|}{\textbf{Auditory QA Test}} \\
\hline \hline
\textbf{Metric} & \textbf{Experts} & \textbf{Crowdworkers} \\\hline
SDA & $0.20\pm0.25$ & $-0.34\pm0.09$ \\\hline
Cohen's $\kappa$  & $0.41\pm0.18$ & $0.00\pm0.07$ \\\hline
Cronbach's $\alpha$  & 0.99 & 0.97 \\ \hline
Krippendorff's $\alpha$   & 0.73 & -0.13 \\\hline \hline
\end{tabular}
\label{tab:reward_table}
\end{center}
\end{table}

This analysis validates H1 both visually and numerically. Trained experts are likely more reliable annotators---compared to the most reliable MTurk crowdworker annotators---when they are tasked to annotate the objective ground truth of the two tests we introduce in this paper. While this finding is not surprising, it indicates that the QA tools introduced offer useful ways of detecting such experience gap.

\subsubsection{Engagement Annotation} 

Following the comparative analysis of the two annotator groups on objectively-defined tasks, we compare their reliability to annotate engagement. We focus on SDA values below, as it specifically designed for this task of (relative) time-continuous annotation and is aligned anyway with other reliability metrics (except perhaps Cronbach's  $\alpha$).
For each of the 30 annotated videos we calculate the SDA value of each annotator from the video's ground truth of engagement. As a reminder, the ground truth of each video for each group of annotators (i.e. expert vs. crowdworkers) is derived as the median engagement trace excluding the annotator that is examined for reliability. The result of this analysis is 150 SDA values per annotator group (i.e. 30 videos times 5 annotators per group).

\begin{figure}[!tb]
\centering
\includegraphics[width=\columnwidth]{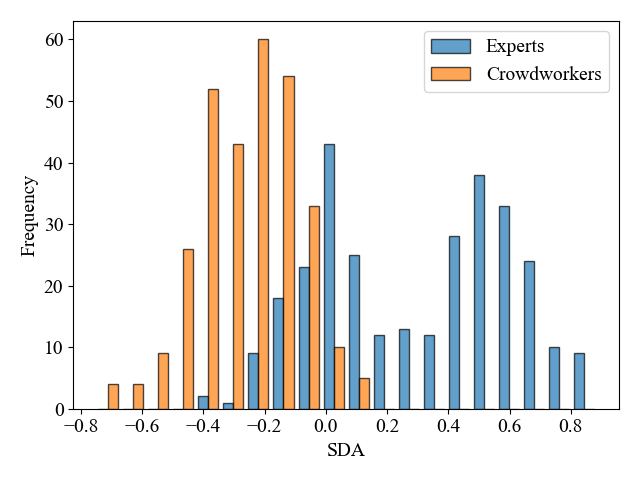}
\caption{Histogram of SDA values for the expert and crowdworker annotators measured across both sessions of engagement annotation tasks.}
\label{fig:engagement_traces}
\end{figure}

Figure \ref{fig:engagement_traces} illustrates the distribution of SDA values for the two annotator groups. Clearly, the expert group manages to annotate engagement with higher degrees of reliability on average. Specifically, the expert group of annotators yield higher agreement with their gold standard signal ($0.304\pm0.196$) compared to the crowdworker group ($-0.26\pm0.04$). A paired-sample t-test between the two distributions reveals that the difference of means between the two groups is significant ($p<0.05$).
Similar results are obtained with the other reliability measures (Cohen's $\kappa$, Krippendorff's $\alpha$)
and are omitted for space considerations. Our findings from this statistical analysis validate H1 once more; this time for affect annotation tasks.

\subsection{H2: Necessity of the Two QA Tests}\label{sec:h2}
A core hypothesis when designing the QA tasks was that their combination covers
both major types of audiovisual stimuli during the annotation process (see Section \ref{sec:qa_auditory}). Since annotators may not pay attention to visual stimuli (e.g. due to small or inappropriate monitors) or audio stimuli (e.g. due to muted or problematic speakers), both a visual QA test and an auditory QA test is necessary. From this analysis, we rephrase H2 as follows: we expect that not all participants that perform reliably in the visual QA task perform similarly in the auditory QA task and vice versa. To test this, we measure the Pearson's correlation ($\rho$) between each annotator's SDA score in the visual QA task with the same annotator's SDA score in the auditory task. For all expert annotators, the correlation between reliability in either QA task is high ($\rho=0.89$) and statistically significant: this means that participants which were good at one task were also good at the other task. This is not surprising since participants were given the same equipment, had experience in such annotation tasks, and we expect that they were attentive to the task due to the laboratory setting in which data was collected. On the other hand, we note a negative correlation between visual QA task reliability and auditory QA task reliability for crowdworkers ($\rho=-0.29$) although the correlation is not statistically significant. While both tasks had low SDA scores for crowdworkers on average (see Table \ref{tab:reward_table}), the low correlation found above indicates that some crowdworkers may be more attentive to visual stimuli than audio stimuli (perhaps due to having a sub-par audio setup or listening to music as they work) and vice versa. While perhaps one QA test would suffice for annotation tasks in controlled settings (due to high correlations between QA tests among experts), the above analysis validates H2: having two QA tests (one for visual attention and one for aural attention) is necessary when dealing with in-the-wild uncontrolled conditions---such as crowdwork---since each QA test returns very different reliability scores for the same user.

\subsection{H3: Predicting an Annotator's Reliability} \label{sec:baselinevsaffect}

\begin{figure}
\centering
\includegraphics[width=\columnwidth]{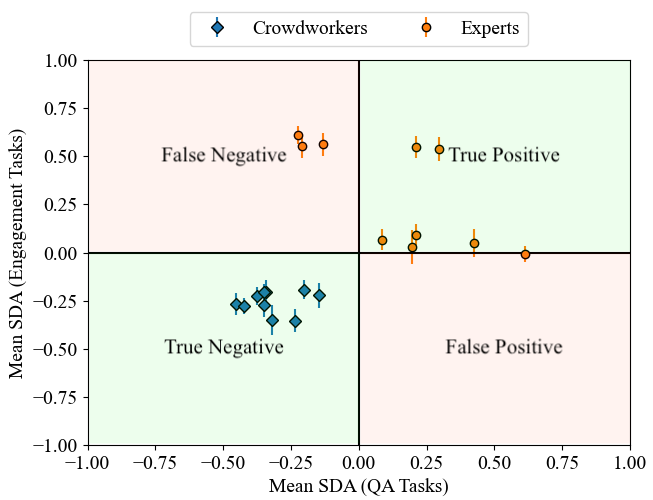}
\caption{Scatter plot of annotators' mean SDA values on the QA tests versus their corresponding mean SDA values across the engagement tasks (including 95\% confidence intervals). We use the class split criterion of 0 (i.e. 50\% agreement) on the mean SDA value of the QA tests to classify between reliable and unreliable annotators. The green quadrants contain annotators the tool correctly classified as reliable or unreliable affect annotators.}
\label{fig:scatter}
\end{figure}

For the purposes of testing H3 we build on certain assumptions regarding SDA scores and identify negative scores as \textit{unreliable}. For each participant, we calculate the mean SDA of the visual and auditory QA tasks. We consider an annotator unreliable if the mean SDA on these two QA tasks is below 0. To validate H3, we test whether this criterion suffices to identify unreliable annotators for the engagement annotation use-case. To assess unreliable annotators for engagement, we calculate the mean SDA score for all 30 annotation tasks that this participant performed. Importantly, for each gameplay video we calculate this participant's SDA from the golden standard which is derived without this participant's trace to avoid data leakage (i.e. the leave-one-participant out method as described in Section \ref{sec:results_groundtruth}). We consider a participant unreliable in the engagement annotation task if their mean SDA score across these 30 videos is negative. By converting these SDA scores into reliable/unreliable labels, we derive Fig.~\ref{fig:scatter}. If we use the QA tasks' SDA to predict unreliable annotators in the engagement annotation, we can treat Fig.~\ref{fig:scatter} as a confusion matrix: unreliable annotators in QA tasks that are reliable in engagement annotation tasks would be false positives, while reliable annotators in QA tasks that are unreliable in engagement annotation tasks would be false negatives. It is evident that all crowdworkers are unreliable both in the QA tasks and the engagement tasks; all 10 crowdworkers are true negatives. Among expert annotators, three are unreliable in QA tasks (all in Session 2) while remaining reliable in the engagement QA task (false negatives). Only one false positive exists among experts, which has a mean SDA on the engagement task marginally below the threshold. This amounts to a correct classification of 16 out of 20 annotators in total (80\% accuracy). While H3 is difficult to validate experimentally (see also Section \ref{sec:discussion}), the above analysis indicates that reliability on QA tests can be a good predictor for general reliability in affect annotation tasks.

\section{Discussion} \label{sec:discussion}

This paper introduced two simple tools for assessing the reliability of annotators rapidly. Our findings suggest that, unsurprisingly, a trained group of annotators (with training on affective computing and AI) perform significantly more reliably than the best crowdworkers we could hire out of 25. Results also suggest that both tools introduced have a great predictive capacity assessing an annotator's reliability in affect annotation tasks, with an accuracy of 80\% in this pilot study.

This study is presented as part of a larger affect annotation experiment through which we solicit affect labels for FPS games in vitro (lab) and in vivo (in the wild). The tests introduced will be assessed on more annotators and annotation tasks. We firmly believe, however, that the sample of 20 participants annotating 2 hours of audiovisual content suffices for supporting the findings of this paper. More annotators and annotation tasks would, in turn, provide more reliability data for training larger models via supervised learning. It is also our desire to test the degree to which the tests are robust and generalizable across annotation domains and datasets involving videos of people manifesting emotion individually or in groups as in \cite{kossaifi2019sewa}.

Our analysis has validated all three hypotheses we made in this paper. As H1 suggests we expected that crowdworkers would not be as reliable as experts and this was evidenced both in QA tests and ensuing tasks that relied on a golden standard signal from the respective groups' annotations. We took several steps to ensure the validity of our findings, such as using a leave-one-participant-out to derive a golden standard signal and reliability on an affect annotation task. The validation of H3, however, relies on using certain SDA values (below 0) as labels for annotator reliability. Additional tests in upcoming work should explore appropriate thresholds for filtering unreliable participants, and how gold standard signals are affected by removing all traces originating from such unreliable participants. 

Another limitation of this work is that the specific properties of the tools are not thoroughly examined and tested. In particular we did not experiment with different ground truths of brightness and pitch and various lengths of the annotation task. Observing the behavior of expert annotators vs. crowdworkers, however, showcased (both qualitatively and quantitatively) that the pattern of the objectively-defined ground truth we designed seem to be independent of the reliability of an annotator.

It is worth noting that the QA tests introduced in this paper are available online\footnote{\url{https://github.com/institutedigitalgames/pagan_qa_tool_sources}} along with the ground truth (brightness and audio pitch fluctuations) so that interested researchers can implement them in their tests. The QA tool was implemented through the PAGAN web annotation framework \cite{PAGAN} which allows researchers to create ad-hoc video annotation tasks. The same repository provides instructions on how to re-implement the protocol described in Section \ref{sec:data_collection}. 

\section{Conclusions} \label{sec:conclusion}

In this paper we introduced an annotator assessment toolkit that can determine rapidly the degree to which an annotator is reliable before they are requested to annotate affect. Building upon existing annotation tools and signal processing methods, we integrate two reliability tests composed of objectively defined visual and auditory tasks within the PAGAN framework \cite{PAGAN}. We validate the tests across 20 annotators---10 experts and 10 crowdworkers---and examine their capacity to capture and predict the reliability of an annotator across a series of engagement annotation tasks. Our findings suggest that the expert annotators are superior---in terms of reliability across four different measures---compared to the best crowdworkers we could hire. Importantly, the introduced assessment tests are capable of predicting the reliability of affect annotators with 80\% accuracy, and thereby constitute a valuable tool for minimizing the resources required for building maximally reliable affect corpora.

\section*{Ethical Impact Statement}

This paper contains a dataset of affect annotations collected from participants in the lab, and through crowd-sourcing on Amazon's MTurk platform. Care was taken to ensure participants gave informed consent about the data collection process, and any personally identifiable information was not stored in the dataset. The dataset contains no potentially offensive data, and followed our University's ethics procedures. The dataset and the tools built will also be made available to the public for other potential studies and scientific reproducibility. To the best of our knowledge, there is no significant potential for the development of negative or deceptive applications using our work, and will not exacerbate existing privacy or discriminatory issues. The extent to which the results of this paper are generalizable across multiple demographics is a topic for future work as we expand on this dataset. The data collection pipeline and analysis did not require the use of any significant compute resources, both in terms of compute power and time.  

\bibliographystyle{IEEEtran}
\bibliography{bibliography}

\end{document}